\documentclass[aoas,MSNbibl,nameyear,dvips]{arximspdf}
\usepackage{dcolumn}
\usepackage{graphicx}

\doi{10.1214/13-AOAS693} 
\volume{8}
\issue{1}
\pubyear{2014}
\firstpage{430}
\lastpage{447}

\makeatletter
\newcolumntype{d}[1]{D{.}{.}{#1}}
\makeatother

\begin{document}
\begin{frontmatter}

\title{A time-varying shared frailty model with application to
infectious
diseases\thanksref{T1}}
\runtitle{A time-varying shared frailty model}
\thankstext{T1}{Supported by funding from the Medical Research
Council and a Royal Society Wolfson Research Merit Award to C.~P. Farrington.}

\begin{aug}
\author[a]{\fnms{Doyo G.} \snm{Enki}\ead[label=e1]{Doyo.Enki@open.ac.uk}},
\author[a]{\fnms{Angela} \snm{Noufaily}\ead[label=e2]{Angela.Noufaily@open.ac.uk}}
\and
\author[a]{\fnms{C. Paddy} \snm{Farrington}\corref{}\ead
[label=e3]{Paddy.Farrington@open.ac.uk}}
\runauthor{D.~G. Enki, A. Noufaily and C.~P. Farrington}
\affiliation{Open University}
\address[a]{Department of Mathematics and Statistics\\
Open University\\
Walton Hall\\
Milton Keynes MK7 6AA\\
United Kingdom\\
\printead{e1}\\
\phantom{E-mail:\ }\printead*{e2}\\
\phantom{E-mail:\ }\printead*{e3}} 
\end{aug}

\received{\smonth{1} \syear{2013}}
\revised{\smonth{9} \syear{2013}}

%
\begin{abstract}
We propose a new parametric time-varying shared frailty model to
represent changes over time in population heterogeneity, for use with
bivariate current status data. The model uses a power transformation of
a time-invariant frailty $U$, and is particularly convenient when $U$
is a member of the generalized gamma family. This model avoids some
shortcomings of a previously suggested time-varying frailty model,
notably time-dependent support. We describe some key properties of the
model, including its relative frailty variance function in different
settings and how the model can be fitted to data. We describe several
applications to shared frailty modeling of bivariate current status
data on infectious diseases, in which the frailty represents
age-dependent heterogeneity in contact rates or susceptibility to infection.
\end{abstract}



\end{frontmatter}
%
\section{Introduction}\label{sec1}

A standard way of representing individual heterogeneity in the hazard
rate of an event of interest is through the multiplicative frailty model
\[
\lambda(t,U) = U \lambda_0(t),
\]
where $U$ is a positive random variable, the frailty, $\lambda(t,U)$ is
the hazard at time~$t$ of an individual with frailty $U$, and $\lambda
_0(t)$ is a baseline hazard common to all individuals in the population
[\citet{AalBorGje08}, \citet{DucJan08},
\citet{Wie11}]. The degree of heterogeneity of the population is then
characterized by the variance of $U$.

In certain circumstances, one may be interested in how the
heterogeneity of the population might vary over time as a result of
changes in individuals' frailties. Such variation might occur in
medical applications, for example, resulting from changes in
individuals' health or behavior. The motivation for this work,
revisited later in the paper, stems from the need to incorporate
unmeasured heterogeneity in contact rates between individuals when
estimating hazards of infection from samples of serological data. Such
heterogeneities are likely to evolve with age, owing to changes in behavior.

In a single sample of data, it is not possible mathematically to
disentangle the baseline survivor function from the frailty
distribution. For this reason, we specifically restrict attention to
shared frailty models, in which this particular identifiability problem
does not occur. This setting is very natural for our intended
application to serological survey data, which arise very commonly in
practice and are often the main primary source of data for infectious
disease modeling. Since serum samples taken from a collection of
individuals are usually tested for several different infections, the
data are typically multivariate and a shared frailty framework arises
very naturally. New identifiability issues ensue, however, which will
be discussed later in the paper.

For simplicity of presentation, the model and its properties are
described first in a univariate context. Incorporating time variation
in the population heterogeneity suggests the more general frailty model
\[
\lambda(t,U) = U(t) \lambda_0(t),
\]
where $U(t)$ is a positive random variable of mean 1 describing how an
individual's frailty evolves over time. To clarify the issues, suppose
that the event of interest is nonterminal. At time $t$, the population
includes people who have experienced the event and people who have not
had the event (the survivors). The unconditional variance of $U(t)$ at
time $t$, var$\{U(t)\}$, describes the heterogeneity of the frailty
$U(t)$ at time $t$ in the entire population. The unconditional
heterogeneity is distinct from the heterogeneity of $U(t)$ in the
population of survivors, described by the conditional variance var$\{
U(t)|T>t\}$.

The relative change over time in the heterogeneity of the survivor
population involves both the change in the frailty variance and also
the selection effect of survival to time $t$. This is quantified by the
relative frailty variance [\citet{FarUnkAna12}],
\[
\operatorname{RFV}^{*}(t) = \frac{\operatorname{var}\{U(t)|T>t\}}{[E\{U(t)|T>t\}]^2}.
\]
In a shared frailty model, an empirical estimate of $\operatorname{RFV}^{*}(t)$ or a
related quantity may be obtained, which can guide the choice of $U(t)$
[Farrington et al. (\citeyear{Faretal13}), Unkel and Farrington (\citeyear{UnkFar12})]. A natural and
flexible framework for representing time-varying frailties is to take
$U(t)$ to be a dynamically evolving stochastic process, such as a
multiplicative Wiener process or a Levy process [Aalen, Borgan and
Gjessing (\citeyear{AalBorGje08})].
However, the undoubted attractiveness of this framework is tempered by
the complexity of the models involved and, more prosaically, by
inherent problems of identifiability when only a single observation is
available on each individual, as is often the case in applications.

These considerations led to the development of a simpler class of
time-varying frailty models of the form
\[
U(t) = w(t,U_1,\ldots,U_k),
\]
where $w(\cdot)$ is a deterministic function of unit mean and $U_1,\ldots
,U_k$ are independent time-invariant frailties. In these models, the
time-invariant frailties are modulated over time, the modulation
occurring in the same way for all individuals in the population. This
model will be appropriate when the evolution of individual frailties is
to some extent governed by common factors, or at least when interest
resides in such an average trajectory. This class of models includes,
for example, piecewise constant frailty models [\citet{PaiTsaOtt94}], for which
\[
w(t, U_1,\ldots,U_k) =\sum_{i=1}^kU_iI_{A_i}(t)=
\prod_{i=1}^k U_i^{I_{A_i}(t)},
\]
where $I_A(t)$ is 1 if $t \in A$ and 0 otherwise, and the $A_i$
partition the positive half-line. Further simplification comes from
restricting these models to the additive or multiplicative forms
\begin{eqnarray*}
w(t,U_1,\ldots,U_k) &=& \sum_{i=1}^k
p_i w(t,U_i),
\\
w(t,U_1,\ldots,U_k)& =& \prod_{i=1}^k
w(t,U_i),
\end{eqnarray*}
where the $w(t,U_i)$ have unit mean and $p_1+\cdots+p_k=1$. In
particular, models with
\[
w(t,U) = 1+(U-1)h(t),
\]
where $0\leq h(t) \leq1$ were introduced in Farrington et al. (\citeyear{Faretal13}).
A detailed discussion of their application to infectious disease data
may be found in Unkel et al. (\citeyear{Unketal}).

A shortcoming of this model is that the range of $U(t)$ is
time-dependent, namely,
\[
1-h(t) < U(t) < \infty,
\]
this restriction being required to maintain $U(t) > 0$. Restricting the
range in this way is artificial and unsatisfactory. Note also that
there is no obvious family of distributions of $U$ on $(0, \infty)$
which is closed under the transformation $1+(U-1)h(t)$ (for given $t$).

In this paper, we propose a new family of time-varying frailty models
which overcomes these shortcomings. In the next section we introduce
the new model. In Section~\ref{sec3} we study the unconditional variance of
$U(t)$ and its relative frailty variance function, and discuss
identifiability issues arising in this type of model. In Section~\ref{sec4} we
describe how to fit a shared frailty model with this time-varying
frailty $U(t)$ to current status data. The performance of the methods
are studied by simulation in Section~\ref{sec5}. Then in Section~\ref{sec6} we apply
these methods to two serological survey data sets.

\section{A new family of time-varying frailty models}\label{sec2}
Our proposal is to replace the linear (in $U$) transformation $w(t,U) =
1+(U-1)h(t)$ by a power transformation, in which
\[
w(t,U) = U^{h(t)}=e^{h(t)\log(U)},
\]
where $U > 0$, $h(t) > 0$ and $h(0)=1$. Note that the range of $U(t) =
w(t,U)$ is $(0,\infty)$ whatever the choice of $h(t)$. Furthermore, if
$U$ belongs to the generalized gamma family with parameters $\theta, k,
\beta>0$ and density
\[
f(u) = \frac{\beta}{\theta^{k\beta}\Gamma(k)}u^{k\beta-1}e^{- (
{u}/{\theta} )^\beta}, \qquad u>0,
\]
where $\Gamma(k)$ is the gamma function, then $U(t)$ is a generalized
gamma with parameters $\theta_t, k, \beta_t>0$ and density
\[
f_t(u) = \frac{\beta_t}{{\theta_t}^{k\beta_t}\Gamma(k)}u^{k\beta_t-1}e^{-
({u}/{\theta_t} )^{\beta_t}},\qquad  u>0,
\]
where $\theta_t=\theta^{h(t)}$ and $\beta_t=\beta/h(t)$. Note that
$\theta_0=\theta$ and $\beta_0=\beta$. This is the parameterization of
the generalized gamma used by \citet{NouJon}.

The generalized gamma family includes the gamma (for $\beta=1$), the
Weibull (for $k=1$) and, as a limiting case for $k\rightarrow\infty$,
the lognormal densities. The generalized gamma distribution has been
used as a frailty density by \citet{BalPen06}. The function
$h(t)$ can be used to denote a smooth transition from one member of the
family to another: for example, $h(t) = e^{-\rho t}+(1-e^{-\rho t})
\beta$, with $\rho> 0$, denotes a transition toward a gamma density as
$t \rightarrow\infty$. Further properties of the family are described
in Cox et al. (\citeyear{Coxetal07}).

More general models may then be built up multiplicatively from such
building blocks, with
\[
U(t) = w(t,U_1,\ldots,U_k) = \prod
_{i=1}^kU_i^{h_i(t)} = \exp \bigl
\{h_1(t) \log(U_1)+\cdots+h_k(t)
\log(U_k) \bigr\}.
\]
Thus,
\[
\log \bigl\{U(t) \bigr\} = \sum_{i=1}^kh_i(t)
\log(U_i).
\]
In general, such models do not belong to the generalized gamma family,
with one exception: if the $U_i(t)$ are lognormal, then so is $U(t)$.
Note that models involving several function $h_i(t)$ may present
identifiability problems and should be used sparingly. An
application-driven example is given in Section~\ref{sec6.2}.

\section{Representing time-varying heterogeneity}\label{sec3}

The frailty $U(t)$ is used to represent individual heterogeneity in
factors impinging upon the event hazard at time~$t$, the degree of
heterogeneity being quantified by its variance. Both unconditional and
conditional variances are of interest, with different interpretations.

\subsection{Unconditional variance}\label{sec3.1}

The moments of $U(t)=U^{h(t)}$ are
\[
E \bigl\{U(t)^r \bigr\} = {\theta_t}^r
\frac{\Gamma(k+r/{\beta_t})}{\Gamma(k)}.
\]
The squared coefficient of variation of $U(t)$ is
\[
\operatorname{CV} \bigl\{U(t) \bigr\} = \frac{\Gamma(k+2/\beta_t)\Gamma(k)}{{\Gamma(k+1/\beta
_t)}^2} - 1.
\]
The derivative of $\operatorname{CV}\{U(t)\}$ with respect to $t$ is
\[
\operatorname{CV}' \bigl\{U(t) \bigr\} = \frac{2}{\beta_0}h'(t) \bigl
\{1+\operatorname{CV} \bigl\{U(t) \bigr\} \bigr\} \bigl\{\psi(k+2/\beta _t)-\psi(k+1/
\beta_t) \bigr\},
\]
where $\psi(x)$ is the digamma function. Since $\psi(x)$ is increasing
on $\mathbb{R}^+$, $h(t)$ and $\operatorname{CV}\{U(t)\}$ have the same turning
points; if $h(t)$ is monotone decreasing to zero, then so is $\operatorname{CV}\{U(t)\}$.

When the event of interest is not terminal, and mortality can be
ignored, the squared coefficient of variation $\operatorname{CV}\{U(t)\}$ describes
the degree of heterogeneity in the population at time $t$. The
variation in heterogeneity is thus described in qualitative terms by
the function $h(t)$.

In applications, it is convenient to ensure that $U(t)$ has unit mean,
which makes it easier to specify a model for the baseline hazard.
Accordingly, we shall normalize $U(t)$ so that it has unit mean, by
dividing $U^{h(t)}$ by its mean and redefining $U(t)$ as follows:
\[
U(t) = \frac{\Gamma(k)}{\theta_t\Gamma(k+1/{\beta_t})}U^{h(t)}.
\]
The (squared) coefficient of variation is unaffected by this
normalization. It may also be desirable to set $E(U) = 1$. This is
readily achieved by setting
\[
\theta= \frac{\Gamma(k)}{\Gamma(k + 1/\beta)},
\]
thus reducing the number of parameters to be estimated.

\subsection{Relative frailty variance: Time invariant case}\label{sec3.2}

When $h(t)=1$ and the event is not terminal, the heterogeneity in the
population does not vary with $t$, unless $U$ is also associated with
mortality in the population. However, the heterogeneity within the
subpopulation who have not experienced the event (the event survivors)
will vary, owing to selection effects. This is described by the
relative frailty variance, or the conditional squared coefficient of variation
\[
\operatorname{RFV}^{*}(t) = \frac{\operatorname{var} \{U|T>t \}}{ \{E(U|T>t) \}^2}.
\]

This can be scaled so that it does not depend on the baseline hazard;
this scaled version is denoted $\operatorname{RFV}(s)$. In shared frailty models,
$\operatorname{RFV}(s)$ is closely related to the cross-ratio function [\citet{Oak89}],
which can be used to guide the choice of~$U$.

The generalized gamma is a special case of the extended generalized
gamma and inverse Gaussian (Egg) family of distributions, discussed in
Farrington, Unkel and
Anaya-Izquierdo (\citeyear{FarUnkAna12}). Specifically, the Egg family of densities is
\[
f(u;\alpha,\beta,\theta,\lambda) = \frac{1}{I^{*}(\alpha,\beta,\theta,\lambda)}
\biggl(\frac{u}{\theta}
\biggr)^{\alpha-1}e^{-\lambda ({u}/{\theta} )}e^{- (
{u}/{\theta} )^\beta}, \qquad u>0,
\]
where $\alpha,\beta,\theta,\lambda>0$ and
\[
I^{*}(\alpha,\beta,\theta,\lambda)=\int_0^\infty
\biggl(\frac{u}{\theta} \biggr)^{\alpha-1}e^{-\lambda ({u}/{\theta
} )}e^{- ({u}/{\theta} )^\beta}\,\mathrm{d}u.
\]
Hence, the generalized gamma is a member of the Egg family with $\lambda
=0$ and $\alpha= k\beta$. Its scaled relative frailty variance
function is therefore
\[
\operatorname{RFV}(s)=\frac{I^{*}(k\beta+2,\beta,1,s\theta)I^{*}(k\beta,\beta,1,s\theta
)}{I^{*}(k\beta+1,\beta,1,s\theta)^2}-1.
\]
This tends to the limit $(k \beta)^{-1}$ as $s \rightarrow\infty$. It
is monotone decreasing for $0 < \beta< 1$, monotone increasing for
$\beta> 1$, and constant when $\beta= 1$ (in which case the density
reduces to the gamma).

%

\subsection{Relative frailty variance: Time-varying case}\label{sec3.3}

When $h(t)$ is not identically~1, the heterogeneity in event survivors
can also be summarized by the relative frailty variance $\operatorname{RFV}^{*}(t)$,
now defined as
\[
\operatorname{RFV}^{*}(t) = \frac{\operatorname{var} \{U(t)|T>t \}}{[E \{
U(t)|T>t \}]^2}.
\]
However, because there are two time scales involved, namely, that at
which events arise and that at which $h(t)$ changes, this can no longer
be re-expressed in a way that does not depend on the baseline hazard.

For arbitrary $U$ and $h(t)$, no explicit expressions for $\operatorname{RFV}^{*}(t)$
are available. Let
\[
I_j(t) = \int_0^{\infty}
U(t)^j \exp \biggl\{-\int_0^tU(s)
\lambda_0(s) \,\mathrm{d}s \biggr\}f(u)\,\mathrm{d}u,
\]
where $f$ is the density of $u$. Then the relative frailty variance is
\[
\operatorname{RFV}^*(t) = \frac{I_2(t) I_1(t)}{I_1(t)^2} - 1.
\]

\begin{figure}
\includegraphics{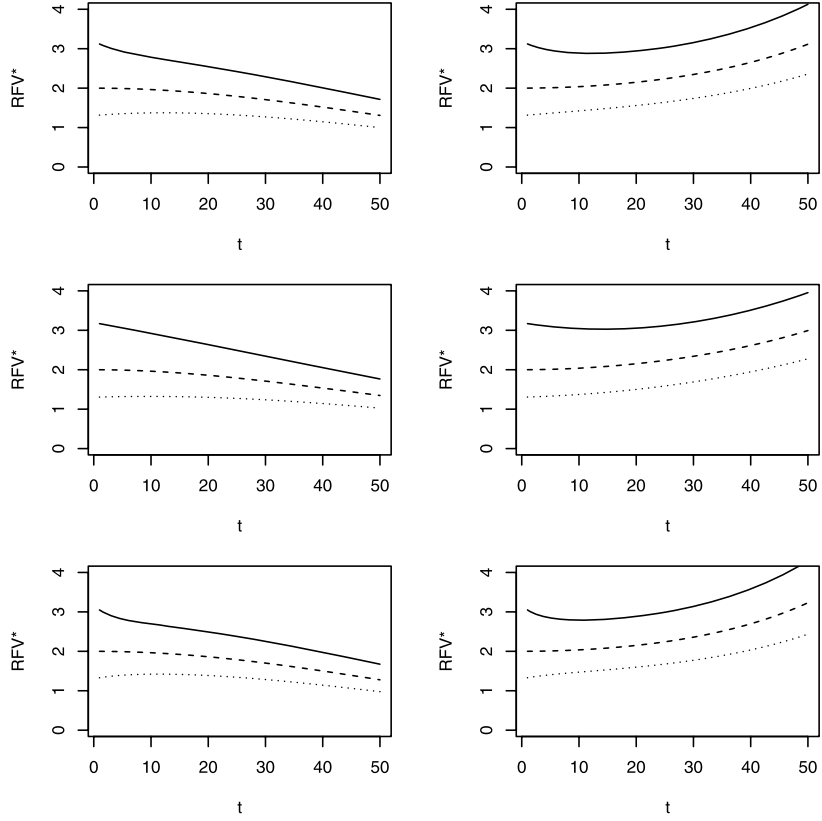}
\caption{Relative frailty variance function $\operatorname{RFV}^{*}(t)$ for
time-varying generalized gamma frailty $U^{h(t)}$ with $E(U) = 1$ and
$h(t) = \exp(-\rho t^2)$. Parameter values are $k = 0.5$ and, top to
bottom in each panel, $\beta= 0.8,1,1.25$. Left panels: $h(t)$
decreasing ($\rho= 0.0001$). Right panels: $h(t)$ increasing ($\rho=
-0.0001$). Top panels: hazard constant, $\lambda(t) = e^{-3.34}$.
Middle panels: hazard increasing, $\lambda(t) = \exp(-4.5+t/25)$.
Bottom panels: hazard decreasing, $\lambda(t)=\exp(-2.5-t/25)$.}
\label{Fig2}
\end{figure}

This can be evaluated numerically. Figure~\ref{Fig2} shows several
plots of $\operatorname{RFV}^{*}(t)$ for $U(t)= U^{h(t)}/E\{U^{h(t)}\}$ with $E(U) =
1$, for different values of $k$ and $\beta$ and contrasting choices of
$h(t)$ and baseline hazard. The baseline hazards are all chosen to have
roughly the same integrated hazard over the range of $t$ displayed. The
plots show that $\operatorname{RFV}^*(t)$ is not greatly influenced by the baseline
hazard, which depends primarily on $h(t)$ and the parameters $k$ and
$\beta$. $\operatorname{RFV}^{*}(t)$ can be estimated empirically in shared frailty
models, and so can be used for inference about $U(t)$.

\subsection{Identifiability issues}\label{sec3.4}

In a shared time-invariant frailty model, it is possible to separate
the baseline hazard functions from the frailty. The function $\operatorname{RFV}(s)$
is then equivalent to the cross-ratio function [\citet{Oak89}]. With
bivariate right-censored data, a plot of $\operatorname{RFV}(s)$ can thus be
obtained
directly. For current status data, the related association measure $\phi$,
which tracks the cross-ratio function, can be obtained [\citet{UnkFar12}]. Briefly, $\phi$ is derived at each sampling time
$t$ from the association parameter for the Clayton copula relating the
empirical marginal and joint survivor functions at time $t$.

When a time-varying frailty $U(t)$ is introduced, it remains possible
to separate the baseline hazard from the frailty, as above, using
$\operatorname{RFV}^{*}(t)$ or the related measure $\phi$. However, it is not possible
to distinguish between time-variation in $U(t)$ from selection effects
reflecting the choice of frailty distributions, and the direct
connection with the cross-ratio function is lost. To clarify the issue,
suppose that the variance of $U$ is small, then $U(t)$ [not normalized,
but parameterized so that $E(U) = 1$] can be approximated linearly to
the first order as
\[
U(t) \simeq1+(U-1)h(t),
\]
which thus approximates the time-varying frailty model described in
Farrington, Unkel and
Anaya-Izquierdo (\citeyear{FarUnkAna12}). Let $\operatorname{RFV}^{*}(t)$ denote the relative frailty
variance of $U(t)$ and $\operatorname{RFV}_0^{*}(t)$ the relative frailty variance of
$U$. Also, let $\mu_c(t)=E[U|T>t]$ be the mean of $U$ in survivors at
time $t$. Direct calculation then yields
\[
\operatorname{RFV}^{*}(t) \simeq \operatorname{RFV}_0^{*}(t) \biggl[
\frac{h(t)}{h(t)+\mu
_c(t)^{-1}(1-h(t))} \biggr]^2.
\]
Note that this expression differs from equation (23) of Farrington, Unkel and
Anaya-Izquierdo (\citeyear{FarUnkAna12}), which contains an error. As noted there, the variation has
two components: a selection effect stemming from $\operatorname{RFV}_0^{*}(t)$ and a
component governed largely, but not exclusively, by $h(t)$. For
example, if $h(t)$ tends to zero, so will this term.

Thus, if $\operatorname{RFV}^{*}(t)$ is observed to change over time, it is not
usually possible to identify whether this is due to changing
heterogeneity, as represented by a nonconstant $h(t)$, or a selection
effect, represented by a nonconstant $\operatorname{RFV}_0^{*}(t)$, or both, without
recourse to external information: the two effects are confounded. This
is unfortunate because distinguishing between selection effects and
evolving heterogeneity can be important in some applications.

Nevertheless, it is possible, and useful, to fit and contrast the two
models corresponding to the most extreme scenarios: gamma $U$ [which
has constant $\operatorname{RFV}_0^{*}(t)$] with time-varying $h(t)$, on the one hand,
and nongamma $U$ [which has nonconstant $\operatorname{RFV}_0^{*}(t)$] with no time
variation, namely, $h(t) = 1$. For the first option, the shape of the
association plot (whether the empirical cross-ratio function or the
empirical plot of $\phi$) can be used to suggest suitable
parameterizations of $h(t)$. In practical applications, including those
described below, it is often found that the selection effects of
available parametric frailty models cannot alone reproduce the observed
patterns. In some circumstances, this is due to theoretical
restrictions on the shapes of $\operatorname{RFV}(s)$ [Farrington, Unkel and
Anaya-Izquierdo (\citeyear{FarUnkAna12})]. More
generally, this suggests either that the available models for
time-invariant frailties are insufficiently flexible or that
time-variation in heterogeneity is the more plausible explanation.

\section{Fitting the model to data}\label{sec4}

Throughout, we shall use
\[
U(t) = \frac{U^{h(t)}}{E\{U^{h(t)}\}},
\]
so that $E\{U(t)\} = 1$. Furthermore, we shall take $E(U) = 1$, so that
the density of~$U$ involves just the two parameters $k$ and $\beta$.
Write $\mu(t) = E\{U^{h(t)}\}$.

\subsection{Evaluating the survivor function}\label{sec4.1}

Fitting the model to data requires the population survivor function
(i.e., the probability of remaining event-free) to be evaluated.
This is
\[
S(t) = E \bigl\{S(t|U=u) \bigr\} = \int_0^\infty
\exp \biggl(-\int_0^t u^{h(s)}
\frac
{\lambda_0(s)}{\mu(s)} \,\mathrm{d}s \biggr) f(u)\,\mathrm{d}u.
\]

In order to avoid evaluating the double integral, an approximate
approach is used. Suppose that time is measured in discrete units of
length $\delta$, in such a way that every time point at which an
observation is made is a multiple of $\delta$. The functions $\lambda
_0(t)$ and $h(t)$, and hence $\mu(t)$, may be approximated by step
functions with steps at the points $m\delta$, $m = 1,2,\ldots,$ taking
the values $\lambda(m\delta)$, $h(m\delta)$ and $\mu(m\delta)$,
respectively, on the interval $((m-1)\delta, m\delta]$. Then, for $t
=j\delta$,
\[
\int_0^t u^{h(s)}\frac{\lambda_0(s)}{\mu(s)}
\,\mathrm{d}s\simeq\delta \sum_{i=1}^{j}
u^{h(i\delta)}\frac{\lambda_0(i\delta)}{\mu(i\delta)}.
\]

The remaining outer integral over $u$ can then be evaluated
numerically; we have used the \texttt{integrate} function in \texttt{R},
version 2.14.0 [\citet{Tea12}]. Convergence problems
may arise when $h(t)$ is increasing; accordingly, we constrained $h(t)$
to be decreasing (or unity) as required in our applications.

In the next section, we will consider a two-component frailty model of
the form
\[
U(t)=U^{h(t)}V,
\]
where $V$ is gamma with mean $1$ and variance $1/k_2$, and density
$g(v)$. Thus,
\begin{eqnarray*}
S(t)&=&\int_0^\infty\int_0^\infty
\biggl\{\operatorname{exp} \biggl(-\int_0^tu^{h(s)}v
\frac{\lambda_0(s)}{\mu(s)}\,\mathrm{d}s \biggr) \biggr\} f(u)g(v)\,\mathrm{d}u\,\mathrm{d}v
\\
&=& \int_0^\infty \biggl[\int
_0^\infty \biggl\{\operatorname{exp} \biggl(-v\int
_0^tu^{h(s)}\frac{\lambda_0(s)}{\mu(s)}\,\mathrm{d}s
\biggr) \biggr\} g(v)\,\mathrm{d}v \biggr] f(u)\,\mathrm{d}u.
\end{eqnarray*}
For $t = j\delta$, consider the expression in the square bracket and write
\begin{eqnarray*}
I(t,u)&=&\int_0^tu^{h(s)}
\frac{\lambda_0(s)}{\mu(s)}\,\mathrm{d}s
\\
& \simeq& \delta\sum_{i=1}^{j}u^{h(j\delta)}
\frac{\lambda_0(j\delta
)}{\mu(j\delta)}.
\end{eqnarray*}
Then, using the Laplace transform for a gamma random variable, we obtain
\[
\int_0^{\infty} e^{-v I(t,u)}g(v)\,\mathrm{d}v =
\biggl[\frac
{k_2}{k_2+I(t,u)} \biggr]^{k_2}
\]
and, hence,
\[
S(t)=\int_0^\infty \biggl[\frac{k_2}{k_2+I(t,u)}
\biggr]^{k_2}f(u)\,\mathrm{d}u.
\]

This last expression can be integrated numerically.

\subsection{Shared frailty model for current status data}\label{sec4.2}

The parameters of the frailty distribution(s) and the function $h(t)$
are readily estimated from a shared frailty model for multivariate
survival data [Aalen, Borgan and
Gjessing (\citeyear{AalBorGje08}), \citet{DucJan08},
\citet{Wie11}]. We restrict attention to the bivariate frailty model
linking two hazard functions with a common frailty:
\[
\lambda_1(t|U,V,\ldots) = U(t)\lambda_{01}(t),\qquad
\lambda_2(t|U,V,\ldots) = U(t)\lambda_{02}(t).
\]

We consider estimation based on bivariate current status data, commonly
available from serological surveys of infectious diseases. At time $t$
we have a 4-tuple $n_{ijt}$ ($i,j=0,1$), where $n_{00t}$ denotes the
number of individuals experiencing neither event by age $t$, $n_{10t}$
the number of individuals experiencing event 1 but not event 2 by time
$t$, and so on. Let $S_{ij}(t)$ denote the corresponding probabilities,
for example, $S_{00}(t)$ is the probability that an individual of age
$t$ has not experienced either event by time $t$. Then
\begin{eqnarray*}
S_{00}(t) &=&\mathrm{E} \biggl\{\operatorname{exp} \biggl(-\int
_0^t U(s) \bigl[\lambda _{01}(s)+
\lambda_{02}(s) \bigr]\,\mathrm{d}s \biggr) \biggr\},
\\
S_{01}(t) &=& \mathrm{E} \biggl\{\operatorname{exp} \biggl(-\int
_0^t U(s)\lambda _{01}(s)\,\mathrm{d}s
\biggr) \biggr\}-S_{00}(t),
\\
S_{10}(t) &=& \mathrm{E} \biggl\{\operatorname{exp} \biggl(-\int
_0^t U(s)\lambda _{02}(s)\,\mathrm{d}s
\biggr) \biggr\}-S_{00}(t),
\\
S_{11}(t) &=& 1-S_{00}(t)-S_{01}(t)-S_{10}(t).
\end{eqnarray*}

These probabilities are evaluated by discretizing the functions $h(t)$,
$\lambda_{01}(t)$ and $\lambda_{02}(t)$ as described above. Let $\gamma
$ denote the vector of parameters describing $f(u)$, $h(t)$, $\lambda
_{01}(t)$ and $\lambda_{02}(t)$. The multinomial log-likelihood kernel
is then
\[
\operatorname{Loglik}(\gamma) = \sum_t\sum
_{i,j=0}^1n_{ijt}\log \bigl(S_{ij}(t)
\bigr). 
\]

This was optimized using function \texttt{nlm} in \texttt{R}, version
2.14.0 [\citet{Tea12}]. Approximate confidence
intervals were obtained using the profile likelihood method. Goodness
of fit was assessed using the deviance, and models were compared using
the AIC.

\section{Simulations}\label{sec5}

We checked the performance of the procedures suggested for current
status data in a small simulation study. The parameter choices for the
simulations broadly reflect the patterns observed in the data to be
analyzed in the next section, namely, declining relative frailty variances.

We assumed constant baseline hazards $\lambda_{01}(t) =\lambda_{02}(t)
= 0.05$, and obtained the survivor functions $S_{ij}(t)$, $i,j = 0,1,$
for two scenarios: (a) $U$ gamma with mean 1 and variance $k^{-1}$ with
exponentially declining heterogeneity $h(t) = \exp(-\rho t^2)$, and (b)
$U$ generalized gamma with mean 1 and parameters $k,\beta$, and
constant heterogeneity ($\rho= 0$). For scenario (b), the model was
parameterized using $\beta$ and $\alpha= k \beta$, to reduce
correlations between the parameter estimates. We generated 4-nomial
samples of size $n=200$ and $n=50$ at each year $t = 1,\ldots,50$. The
procedure was run $N=400$ times for each parameter combination. The
results for scenario (a) are shown in Table~\ref{Tab1}, and those for
scenario (b) in Table~\ref{Tab2}.

\begin{table}%
\caption{Simulation results for gamma frailty with declining
heterogeneity}\label{Tab1}
\begin{tabular*}{\textwidth}{@{\extracolsep{\fill}}lccccc@{}}
\hline
&& $\bolds{\operatorname{Bias} (\hat\rho)}$ & $\bolds{\operatorname{RMSE} (\hat\rho)}$ &
$\bolds{\operatorname{Bias} (\hat k)}$ & $\bolds{\operatorname{RMSE}(\hat k)}$\\
\hline
\multicolumn{6}{c}{$n=200$}\\
$k=1$ & $\rho= 0.01 $ & 0.0007 & 0.0045 & 0.0311 & 0.2959\\
& $\rho= 0.002$ & 0.0000 & 0.0003 & 0.0031 & 0.1045\\
$k=0.2$ & $\rho= 0.01 $ & 0.0001 & 0.0011 & 0.0011 & 0.0228\\
& $\rho= 0.002$ & 0.0000 & 0.0001 & 0.0004 & 0.0114\\[3pt]
\multicolumn{6}{c}{$n=50$}\\
$k=1$ & $\rho= 0.01 $ & 0.0036 & 0.0143 & 0.0906 & 0.6470\\
& $\rho= 0.002$ & 0.0000 & 0.0007 & 0.0178 & 0.2218\\
$k=0.2$ & $\rho= 0.01 $ & 0.0004 & 0.0022 & 0.0004 & 0.0448\\
& $\rho= 0.002$ & 0.0000 & 0.0002 & 0.0024 & 0.0229\\
\hline
\end{tabular*}
\end{table}

\begin{table}[b]
\caption{Simulation results for generalized gamma frailty with constant
heterogeneity}\label{Tab2}
\begin{tabular*}{\textwidth}{@{\extracolsep{\fill}}lccccc@{}}
\hline
&& $\bolds{\operatorname{Bias} (\hat\beta)}$ & $\bolds{\operatorname{RMSE} (\hat\beta)}$ &
$\bolds{\operatorname{Bias} (\hat\alpha)}$ &
$\bolds{\operatorname{RMSE} (\hat\alpha)}$\\
\hline
\multicolumn{6}{c}{$n=200$}\\
$k=1$ & $\beta=0.7$ & 0.0064 & 0.1983 & 0.0402 & 0.1524\\
& $\beta=0.4$ & 0.0214 & 0.1269 & 0.0291 & 0.1647\\
$k=0.2$ & $\beta=0.7$ & 0.0310 & 0.2381 & 0.0042 & 0.0227\\
& $\beta=0.4$ & 0.0087 & 0.1199 & 0.0021 & 0.0129\\[3pt]
\multicolumn{6}{c}{$n=50$}\\
$k=1$ & $\beta=0.7$ & 0.1887 & 0.9342 & 0.1456 & 0.5433\\
& $\beta=0.4$ & 0.1172 & 0.3673 & 0.0797 & 0.4713\\
$k=0.2$ & $\beta=0.7$ & 0.3541 & 1.1469 & 0.0116 & 0.0722\\
& $\beta=0.4$ & 0.0633 & 0.3813 & 0.0022 & 0.0196\\
\hline
\end{tabular*}
\end{table}

When $n=200$, the estimated bias in $\rho$ and $k$, and in $\beta$ and
$\gamma$, is small, seldom exceeding 5\% of the true parameter value.
The estimated root mean squared errors (RMSE) are larger, reflecting
the lack of information available from current status data. When $n =
50$, the biases and RMSE values are greater, owing to the sparseness of
the data at young ages and the consequent lack of information on $\rho$
and $\beta$. The RMSE values suggest that, in scenario (a), larger
values of $\rho$, corresponding to rapid drops in heterogeneity, are
more difficult to estimate, whereas in scenario (b), $\beta$ becomes
more difficult to estimate as it approaches 1, that is, as the
distribution of $U$ becomes closer to the gamma. In all cases, the
baseline hazard parameters are estimated with little bias (not shown).

We obtained asymptotic standard errors for the parameter estimates from
a numerical estimate of the Hessian matrix. The means of these standard
errors were generally less than the standard deviations of the
simulated parameter estimates, and Wald confidence intervals had
coverage probabilities lower than the nominal values (results not
shown). The discrepancy was most marked for the parameters relating to
the frailty ($\rho$, $k$, $\beta$, $\alpha$). We conclude that
asymptotic standard errors may be unreliable in samples of moderate
size, and recommend that interval estimates be obtained by
bootstrapping or profile likelihood.

We also undertook a further simulation to investigate the robustness of
inferences about $\rho$ to misspecification of the frailty
distribution. Thus, we generated data (400 replicates with $n = 200$ at
each year) from a generalized gamma frailty with $\beta= 0.5$ and $k =
2$, with $U(t)$ of unit mean. We fitted the same gamma model as for
Table~\ref{Tab1} to these simulated data. The results are shown in
Table~\ref{Tab1a}. As expected, the bias for $\hat\rho$ is worse than
in Table~\ref{Tab1} (for $n=200$), though only marginally so, and the
RMSE values are little affected. We conclude that the methods are
reasonably robust to mild misspecification of the frailty within the
generalized gamma family.

\begin{table}%
\tablewidth=200pt
\caption{Simulation results for misspecified gamma frailty model}\label{Tab1a}
\begin{tabular*}{200pt}{@{\extracolsep{\fill}}lcc@{}}
\hline
$\bolds{\rho}$& $\bolds{\operatorname{Bias} (\hat\rho)}$ &
$\bolds{\operatorname{RMSE} (\hat\rho)}$ \\
\hline
$0.01 $ & 0.0030 & 0.0044\\
$0.002$ & 0.0006 & 0.0006\\
$0$ & 0.0001 & 0.0001\\
\hline
\end{tabular*}
\end{table}

\section{Applications}\label{sec6}

We illustrate the methods with applications to two contrasting data
sets, each involving a pair of infections. The data are serological
survey data from the UK and are described in detail in \citet{Faretal13}
and \citet{Unketal}. Each individual
of age $t$ is tested by two laboratory assays to determine whether he
or she has been infected or not at some time prior to $t$. The data are
thus paired current status data and are observed at calendar years of
age $t = 1,2,\ldots,M$. Each paired sample contributes a likelihood term
as described in the previous subsection. In some instances, only one of
the test results is available. In this case, the likelihood
contributions are obtained from the corresponding marginal
probabilities. We fit different models for $U(t)$ according to whether
the two infections share a mode of transmission.

\subsection{Parvovirus B19 and Cytomegalovirus infections}\label{sec6.1}

Parvovirus B19 is transmitted via droplets via the respiratory route,
whereas Cytomegalovirus is transmitted by oral ingestion of
contaminated secretions. Thus, the route of transmission is different
for the two infections. In childhood, transmission via these two routes
is likely to be confounded, owing to the closeness of contacts between
young children. Heterogeneity of contacts in early childhood---for
example, owing to variation in nursery attendance---is likely,
therefore, to induce an association between the infections. This is
unlikely to persist into adulthood, since the infections are
transmitted differently.

\begin{figure}

\includegraphics{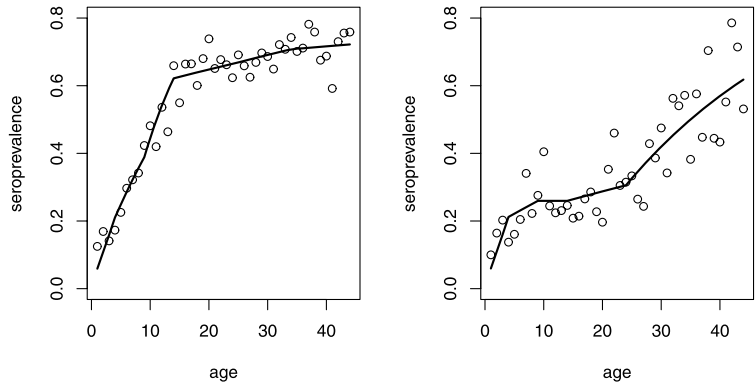}

\caption{Seroprevalence of Parvovirus B19 (left panel) and
Cytomegalovirus (right panel) infections with age (years). The lines
show the fitted values obtained from the model with shared frailty
$U(t) \propto U^{h(t)}$ with gamma $U$ and exponentially declining
$h(t)$.}\label{Fig3}
\end{figure}

\begin{figure}

\includegraphics{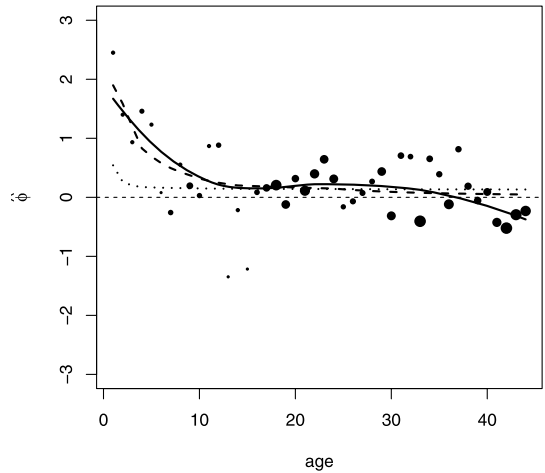}

\caption{Association between Parvovirus B19 and Cytomegalovirus by age
(years). The dots represent the empirical values of the association
parameter $\hat\phi$ (with area inversely proportional to the
empirical variance of $\hat\phi$). The full line is a LOESS curve. The
curved dashed line shows the fitted values obtained from the model
with shared frailty $U(t) \propto U^{h(t)}$ with gamma $U$ and
exponentially declining $h(t)$. The curved dotted line shows the
fitted values obtained from the model with time-invariant generalized
gamma frailty $U$. The horizontal dashed line represents no
association.}\label{Fig4}
\end{figure}

Figure~\ref{Fig3} shows the observed proportions with a positive test
result, or seroprevalences, by age, and Figure~\ref{Fig4} shows the
association between the two infections in each pair, with a LOESS curve
to represent the trend. The measure of association used here, denoted
$\phi$, is described in \citet{UnkFar12}. It tracks the
relative frailty variance $\operatorname{RFV}^*(t)$ and hence the cross-ratio
function, neither of which can be obtained directly from current status
data.

The association plot suggests that there is a high degree of
heterogeneity at early ages, which declines rapidly with age. As
expected, there is evidence of heterogeneity at young ages, possibly
due to heterogeneity of contacts. An alternative explanation, at young
ages, could be variation in development of the immune system. The
decline in heterogeneity may be related to the homogenizing influence
of school attendance and other learned behaviors.

The LOESS curve in Figure~\ref{Fig4} suggests that the time-varying
frailty model $U(t) \propto U^{h(t)}$ with $h(t) = \exp(-\rho t^2)$
might be an appropriate choice; the constant of proportionality is the
normalizing factor to ensure that $U(t)$ has unit mean for all $t$. We
chose $U$ to be a unit mean gamma random variable with variance
$k^{-1}$ (but also tried a generalized gamma). We also fitted the
selection model, according to which the decline in the strength of
association is due entirely to selection effects. In this model, $U$ is
represented by a unit mean generalized gamma random variable with
parameters $k$ and $\beta$, but there is no time-varying effect, so
$\rho= 0$. In each case the baseline hazards were modeled using
piecewise constant functions.

\begin{table}[b]
\caption{Fit to Parvovirus B19 and Cytomegalovirus infection data}\label{Tab3}
\begin{tabular*}{\textwidth}{@{\extracolsep{\fill}}lccccc@{}}
\hline
\textbf{Model} & \textbf{$\bolds{-}$Loglik}& \textbf{Deviance} & \textbf{df} &
\textbf{$\bolds{p}$-value} & \textbf{AIC}\\
\hline
Gamma with trend & 4352.07 &231.76&206&0.105&8732.14\\
Gen. Gamma, no trend & 4357.35 &242.32&206&0.042&8742.70\\
Gamma, no trend & 4359.04 &245.70&207&0.034&8744.09\\
\hline
\end{tabular*}
\end{table}

The results are in Table~\ref{Tab3}. The best fit is achieved by the
time-varying frailty model with gamma $U$. Unsurprisingly, in view of
the identifiability issues discussed earlier, a generalized gamma $U$
gave no improvement over the gamma for this model, though this lack of
improvement does suggest that the model for $h(t)$ is not grossly
misspecified. The selection model gave a moderately worse fit to the
data. The fitted association curves $\hat\phi$ for the two models are
shown in Figure~\ref{Fig4}, and show that the selection model does not
adequately represent the association. Also included in Table~\ref{Tab3}
is a simple gamma shared frailty model, for which the fit is less good.
The results for this model (as in the next example) differ slightly
from those of \citet{Unketal}, as different age groups were used.
We conclude that the best fitting model is the time-varying frailty
model; the observed and fitted seroprevalences are shown in Figure~\ref{Fig3}. The parameters of this fitted frailty model, with 95\% profile
likelihood confidence intervals, are as follows: $k_1 = 0.168$
$(0.0630, 0.775)$, $\rho= 0.134$ $(0.0272,0.645)$.

\subsection{\textit{Helicobacter pylori} and Toxoplasma infection}\label{sec6.2}

\emph{Helicobacter pylori} and Toxoplasma infection are both
transmitted by the oral route via ingestion of contaminated matter.
Because the infections share a common route of transmission, we might
expect that heterogeneities in adult behavior will be reflected in a
persistent association between the infections.

\begin{figure}

\includegraphics{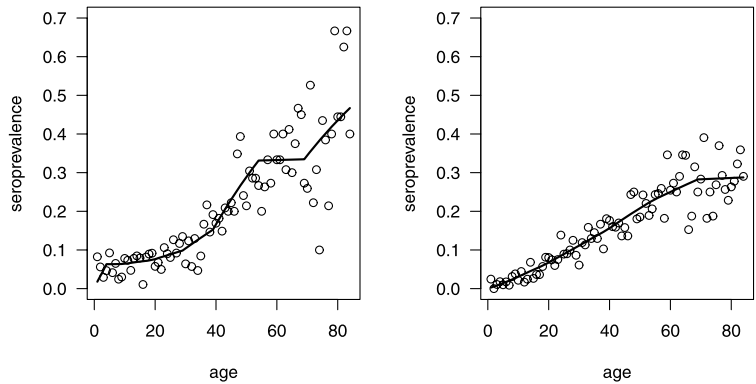}

\caption{Seroprevalence of \textit{Helicobacter pylori} (left panel)
and Toxoplasma (right panel) infections with age (years). The lines
show the fitted values obtained from the model with shared frailty
$U(t) \propto U^{h(t)}V$ with gamma $U$ and $V$ and exponentially
declining $h(t)$.}\label{Fig5}
\end{figure}

\begin{figure}

\includegraphics{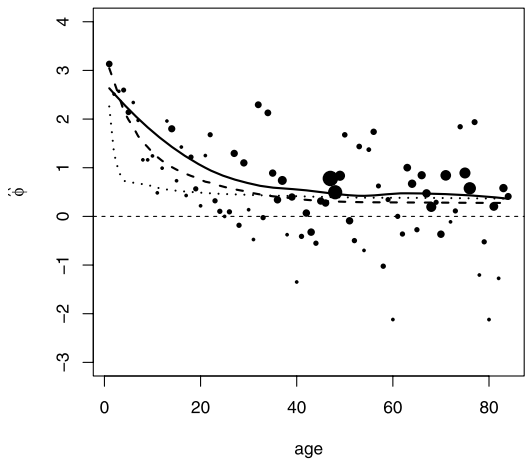}

\caption{Association between \textit{Helicobacter pylori} and
Toxoplasma by age (years). The dots represent the empirical values of
the association parameter $\hat\phi$ (with area inversely proportional
to the empirical variance of $\hat\phi$). The full line is a LOESS
curve. The curved dashed line shows the fitted values obtained from
the model with shared frailty $U(t) \propto U^{h(t)}V$ with gamma $U$
and $V$ and exponentially declining $h(t)$. The curved dotted line
shows the fitted values obtained from the model with time-invariant
generalized gamma frailty $U$. The horizontal dashed line represents
no association.}\label{Fig6}
\end{figure}

Figure~\ref{Fig5} shows the marginal seroprevalences for the two pairs,
and Figure~\ref{Fig6} the association plots. As for Parvovirus B19 and
Cytomegalovirus, there is substantial heterogeneity at young ages,
declining with increasing age. However, the decline is now not to zero:
as expected, there remains a small but persistent association in
adulthood. This is most likely due to heterogeneity in contacts via the
common transmission route, owing to differences in eating habits,
hygiene and environmental factors.

We thus propose a time-varying model for the frailty $U(t)$ involving
two components. The first component is the ``childhood'' component
$U^{h(t)}$ with $h(t) = \exp(-\rho t^2)$, representing heterogeneities
in childhood as before. The second component $V$ represents adult
heterogeneity in behavior associated with transmission by the common
route. Thus, $V$ might represent variation in exposure to the ingestion
of contaminated matter. We assume that $U$ and $V$ are independent
random variables, both of unit mean. The frailty model is thus
\[
U(t) \propto U^{ \exp(-\rho t^2)}V,
\]
the constant of proportionality being the normalizing factor to ensure
$U(t)$ has expectation 1. We assume that both $U$ and $V$ are gamma
with variances $k_1^{-1}$ and $k_2^{-1}$, respectively; we also allowed
$U$ to be generalized gamma. We also fitted a selection model in which
$U(t) = U$ is generalized gamma with parameters $k_1$ and~$\beta$, but
with $\rho= 0$.

The results are in Table~\ref{Tab4}. As for the previous example, for
the time-varying frailty model, allowing a generalized gamma $U$ gave
virtually no improvement over a gamma $U$. The time-varying frailty
model gave only a marginally better fit than the selection model with
constant generalized gamma frailty. However, the observed associations
$\phi$ in Figure~\ref{Fig6} are much more faithfully reproduced by the
time-varying frailty model than by the selection model. Both models fit
better than the simple gamma shared frailty model. We thus select the
two-component time-varying frailty model as the best one; its fit to
the serological data in Figure~\ref{Fig5} is good. The parameters of
this model, with 95\% profile likelihood confidence intervals, are as
follows: $k_1 = 0.0572$ $(0.0215,0.275)$, $k_2 = 3.29$ $(1.87,9.98)$,
and $\rho= 0.0911$ $(0.0502,0.183)$.

\begin{table}
\centering
\caption{Fit to \textit{Helicobacter pylori} and Toxoplasma infection
data}\label{Tab4}
\begin{tabular*}{\textwidth}{@{\extracolsep{\fill}}lccccc@{}}
\hline
\textbf{Model} & $\bolds{-}$\textbf{Loglik}& \textbf{Deviance} &
\textbf{df} & $\bolds{p}$\textbf{-value} & \textbf{AIC}\\
\hline
Gamma/Gamma with trend & 4229.26&399.95&364&0.094&8496.52\\
Gen. Gamma, no trend & 4230.86&403.15&365&0.082&8497.72\\
Gamma, no trend & 4235.09&411.62&366&0.050&8504.18\\
\hline
\end{tabular*}
\end{table}

\section{Final remarks}\label{sec7}
We have presented a simple time-varying frailty model in which
time-invariant frailties are modulated over time by a deterministic
function. The major limitation of this kind of approach is that all
individuals are assumed to share the same trajectories over time. Thus,
it is likely to be applicable only when variation in heterogeneity is
driven by a mechanism common to all individuals, such as ageing.

The present model uses a power function of the (time-invariant)
frailty, rather than a linear function as previously suggested.
Arguably, the new model is more natural and avoids some limitations of
the linear model, notably time-dependent support. However, this benefit
comes at the cost of analytical tractability, which we overcame by a
combination of discretization and numerical integration. Nevertheless,
the new model fits naturally within a broad class of generalized gamma
time-varying frailty models, from which some analytical results may be
exploited.

We focused attention on frailties within the generalized gamma family,
owing to its mathematical tractability and its appropriateness for our
application. This family is reasonably flexible in that it allows for
both monotone increasing and decreasing (and constant) relative frailty
variance functions. However, in some applications, other types of time
trends might be required.

The new model does not provide any resolution of the central conundrum
of time-varying frailty models, namely, how to distinguish between a
selection effect and genuine temporal variation in heterogeneity.
However, it provides some new tools to explore these contrasting
interpretations in a shared frailty context. The applications to
serological survey data reinforce the value of plotting the empirical
and fitted values of the association measure $\phi$. Such plots enable
a more sensitive assessment of model fit than is possible by marginal
observed and expected plots or single numerical summaries of goodness
of fit.

\section*{Acknowledgment}
We thank Richard Pebody (Public Health England, London) for permission
to use the serological data.

%



\printaddresses

\end{document}